\documentclass[fleqn,twoside]{article}
\usepackage{espcrc2}

 \usepackage{latexsym}
 \usepackage{amssymb}

\newcommand{\be}{\begin{equation}}
\newcommand{\ee}{\end{equation}}
\newcommand{\nn}{\nonumber}
\newcommand{\bea}{\begin{eqnarray}}
\newcommand{\eea}{\end{eqnarray}}

\date{May 26, 2003}
\author{C.~Carimalo\address{LPNHE, IN2P3-CNRS, Universit\'e Paris VI,
  F-75252 Paris, France},
  A.~Schiller\address{Institut f\"ur Theoretische Physik and NTZ, Universit\"at Leipzig,
  D-04109 Leipzig, Germany},
 V.G.~Serbo\address{Novosibirsk State
University, 630090, Novosibirsk, Russia}\thanks{This work is
supported in part by INTAS (00-00679), NSh (2339,2003.2) and by
RFBR (03-02-17734). }
\thanks{Contribution presented by V.G. Serbo
at PHOTON 2003, Frascati, Italy} }

\title{
\thispagestyle{empty}
\vspace{-17mm}
\rightline
{\small LU-ITP 2003/012, LPNHE 2003-06~}
\rightline{\small  26 May, 2003~}
\vspace{5mm}
A new method for calculating jet-like QED processes}

\begin{document}

\begin{abstract}
We consider inelastic QED processes, the cross sections of which
do not drop with increasing energy. Such reactions have the form
of two-jet processes with the exchange of a virtual photon in the
$t$-channel. We consider them in the region of small scattering
angles $ m/E \lesssim \theta \ll 1$, which yields the dominant
contribution to their cross sections. A new effective method is
presented to calculate the corresponding helicity amplitudes. Its
basic idea consists in replacing spinor structures for real and
weakly virtual intermediate leptons by simple transition vertices
for real leptons.  The obtained compact amplitudes are
particularly suitable for numerical calculations in jet-like
kinematics.
 \vspace{1pc}
\end{abstract}

\maketitle

\section{INTRODUCTION}

{\bf The subject.} Accelerators with high-energy colliding
$e^+e^-$, $\gamma e$, $\gamma \gamma$ and $\mu^+ \mu^-$ beams are
now widely used or designed to study fundamental interactions.
Some processes of quantum electrodynamics (QED) might play an
important role at these colliders, especially those inelastic
processes the cross sections of which do not drop with increasing
energy. For this reason and since, in principle, the planned
colliders will be able to work with polarized particles, these QED
processes are required to be described in full detail, including
the calculation of their amplitudes with definite helicities of
all initial and final particles.

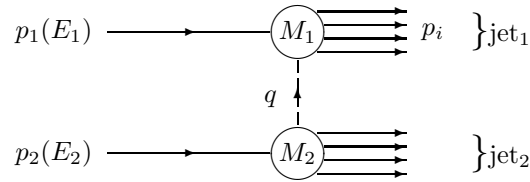
\begin{figure}[!htb]
  \centering
  \unitlength=1.8mm
  \begin{picture}(38.00,14.00)
    \put(23.00,12.00){\circle{4.00}}
    \put(23.00,12.00){\makebox(0,0)[cc]{$M_1$}}
    \put(23.00,3.00){\circle{4.00}}
    \put(23.00,3.00){\makebox(0,0)[cc]{$M_2$}}
    \put(9.00,3.00){\vector(1,0){6.50}}
    \put(9.00,12.00){\vector(1,0){6.50}}
    \put(14.00,3.00){\line(1,0){6.90}}
    \put(14.00,12.00){\line(1,0){6.90}}
    \put(25.00,12.50){\vector(1,0){6.00}}
    \put(25.00,11.50){\vector(1,0){6.00}}
    \put(24.50,13.50){\vector(1,0){6.50}}
    \put(23.00,5.10){\line(0,1){0.90}}
    \put(23.00,6.50){\line(0,1){2.00}}
    \put(23.00,6.50){\vector(0,1){1.40}}
    \put(23.00,9.00){\line(0,1){0.90}}
    \put(24.50,10.50){\vector(1,0){6.50}}
    \put(25.00,3.50){\vector(1,0){6.00}}
    \put(25.00,2.50){\vector(1,0){6.00}}
    \put(24.50,4.50){\vector(1,0){6.50}}
    \put(24.50,1.50){\vector(1,0){6.50}}
    \put(33.00,12.00){\makebox(0,0)[cc]{$p_i$}}
    \put(21.00,7.00){\makebox(0,0)[cc]{$q$}}
    \put(5.00,12.00){\makebox(0,0)[cc]{$p_1(E_1)$}}
    \put(5.00,3.00){\makebox(0,0)[cc]{$p_2(E_2)$}}
    \put(38.00,12.00){\makebox(0,0)[cc]{{\Large\}}jet$_1$}}
    \put(38.00,3.00){\makebox(0,0)[cc]{{\Large\}}jet$_2$}}
  \end{picture}
  \caption{ Generic block diagram of the two-jet process  $e e \to
  {\mathrm {jet}}_1 \ {\mathrm{jet}}_2$.}
  \label{fig:1}
\end{figure}

These reactions have the form of
a two-jet process with the exchange of a virtual photon
$\gamma^*$ in the $t$-channel (Fig.~\ref{fig:1}).
The subject of our consideration (for more details see
Refs.~\cite{CSS1,CSS2} and references therein) is that process at
 high energies ($m_i$ is a lepton mass)
 \be
  s=2p_1 p_2 =4E_1 E_2 \gg m_i^2
  \label{1}
 \ee
for arbitrary  helicities of leptons $\lambda_i=\pm 1/2$ and
photons $\Lambda_i=\pm 1$. The emission and scattering angles
$\theta_i$ are small:
 \be
  \frac{m_i}{E_i} \lesssim \theta_i  \ll 1\,, \;\; m_i \lesssim
  |{\bf p}_{i\perp}| \ll E_i \,.
  \label{2}
 \ee
The corresponding Feynman diagrams of the discussed process up to
$e^5$ order are given in~\cite{CSS1,CSS2}. We mention here only
such processes as: single and double bremsstrahlung and single
pair production in $e^+e^-$ collisions, single pair production in
$\gamma e$ and double pair production in $\gamma \gamma$
collisions.

The processes under consideration have large total cross sections.
Therefore, they present an essential background and they determine
particle losses in the beams and the beam life time. Since all
these reactions can be calculated with high accuracy independently
of any model of strong interaction, they can usefully serve for
monitoring the luminosity and polarization of colliding beams.
Besides, the methods for calculating helicity amplitudes of these
QED reactions can be easily translated to several semihard QCD
processes such as $\gamma \gamma \to q \bar{q} Q \bar{Q}$   and
$\gamma \gamma \to M M'$, $\gamma \gamma \to M q \bar{q}$
(see~\cite{kuraevnp85,mesons}).

All these properties of the jet-like QED processes justify the
growing interest in them from both the experimental and
theoretical communities in high-energy physics. Particular
problems related to these processes were discussed in a number of
original papers and in reviews such as \cite{BGMS,BFKK}. But only
recently (see Ref.~\cite{KSSSh00}) the highly accurate analytical
calculation of the helicity amplitudes of all jet-like processes
up to $e^4$ was completed. In the above-mentioned original papers
different approaches have been used. In our
papers~\cite{CSS1,CSS2} we have developed a new simple and
effective method to calculate those processes. For simplicity, in
this contribution we restrict ourselves to bremsstrahlung
processes only.

{\bf The form of ``the final result''.} At high energies the
region of scattering angles (\ref{2}) gives the dominant
contribution to the cross sections of all QED jet-like processes.
In this region we obtain all helicity amplitudes with high
accuracy, omitting only terms of the order of
 \be
  \frac{m_i^2}{E_i^2} \ , \ \ \theta_i^2 \ , \ \
  \theta_i\cdot\frac{m_i}{E_i}
  \label{3}
 \ee
or smaller. The amplitude $M_{fi}$ has a simple factorized form
\be
  M_{fi}= \frac{s}{q^2} J_1 J_2
  \label{4}
 \ee
where the {\it impact factors} $J_1$  and $J_2$ do not depend on
$s$.  We give  analytical expressions for $J_1$  and $J_2$. They
are not only compact but are also very convenient for numerical
calculations, since large compensating terms are already cancelled.
It is well known that this problem of large compensating terms is
very difficult to manage in all computer packages. The discussed
approximation differs considerably from the known approach of the
CALCUL group in which terms of the order of $m_i/|{\bf
p}_{i\perp}|$ are neglected.

{\bf Three basic ideas:} {\bf (i)} a convenient decomposition of
all 4-momenta into large and small components (using the so-called
Sudakov or light-cone variables); {\bf (ii)} gauge invariance of
the amplitudes is used in order to combine large terms into finite
expressions; {\bf (iii)} the calculations are considerably
simplified in replacing the numerators of lepton propagators by
vertices involving real leptons.

All these ideas are not new. In particular, the last one is the
basis of the equivalent-electron
approximation~\cite{Kessler60,BFKhquasi} and has been used to
calculate some QCD amplitudes with massless quarks. However, the
combination of these ideas leads to a very efficient way in
calculating the amplitudes of interest just in the jet kinematics.

\section{METHOD OF CALCULATION}

{\bf Sudakov or light-cone variables.} We use light-like
4-vectors $P_1$ and $P_2$:
 \bea
\lefteqn{  P_1 = p_1 -{m_1^2\over s }\,\; p_2,\;\; P_2=
  p_2 -{m_2^2\over s }\, p_1\,,}
  \nn\\
\lefteqn{  P_1^2=  {P}_2^2 = 0\,,}
 \eea
and decompose any 4--vector $A$ as
  \bea
\lefteqn{A= x_A P_1 + y_A P_2+ A_\perp\,,\;\;
 A^2 =  s x_A y_A + A_\perp^2 \,,}
  \nn
\\
\lefteqn{x_A=\frac{ 2 A P_2}
  { s}\,,\;\;y_A=\frac{ 2 A P_1} { s} \,.}
  \label{6}
 \eea
The parameters $x_A$ and $y_A$ are the so-called {\it Sudakov
variables}. The 4-vectors $p_i$ of particles from the first jet
have large components along $P_1$ and small ones along $P_2$, i.e.
 \be
x_i =\frac{2p_i P_2}{s} = \frac{E_i}{E_1}\,, \;\; y_i =\frac{2p_i
P_1}{s} = \frac{m^2_i +{\bf p}_{i\perp}^2}
  {s x_i}\,.
  \label{9}
 \ee
Therefore, in the limit $s\to \infty$ the parameters $x_i$ are
finite, whereas $ y_i$ are small. The Sudakov variable $x_i$ is
the fraction of energy of the first incoming particle carried by
the $i$-th final particle. The Sudakov parameters $x_q$ and $y_q$
for the virtual photon are small.

Let $e \equiv e^{(\Lambda)}(k)$ be the polarization 4-vector of
the final photon in the first jet. Using gauge invariance, this
vector can be presented in the form
 \be
  e=y_e P_2+{e}_\perp\,,\;\;
  y_e =\frac{-2 k_\perp {e}_\perp}{sx_k}\,,\;\;
  x_k = \frac{2kP_2}{ s}\,,
  \label{15}
 \ee
where
 \be
  {e}_\perp\equiv {e}_\perp^{(\Lambda)} = -{\Lambda\over
  \sqrt{2}}\, (0,\,1,\,{\rm i} \Lambda ,\,0) =
  -{e}_\perp^{(-\Lambda)\,*}\,.
  \label{16}
 \ee
Therefore, ${e}_\perp$ {\it does not depend} on the 4-momen\-tum
of the photon $k$ contrary to the polarization vector $e$ itself.

{\bf Factorization of amplitudes.} The amplitude of Fig. 1 can be
written as
 \be
  M_{fi} = M_1^{\mu} \; {g_{\mu \nu} \over q^2} \; M_2^\nu ,
  \label{17}
 \ee
where $M_1^\mu$ and $M_2^\nu$ are the amplitudes of the upper and
lower block in Fig.~\ref{fig:1}. It is not difficult to show that this
amplitude can be presented in the simple factorized form (\ref{4}) with
  \be
   \label{18}
  J_1 = {\sqrt{2}\over s}
  \, M_1^{\mu} \, P_{2\mu}\,, \quad J_2 = {\sqrt{2}\over s} \,
  M_2^{\nu} \, P_{1\nu} \,.
 \ee
At high energies, the impact factor $J_1$ depends on $x_i,\; {\bf
p}_{i\perp}$ with $i\in \;$ jet$_1$ and on the helicities of the
first particle and of the particles in the first jet. We use
exactly this form for the concrete calculations.

Due to gauge invariance, we have $ M_1^\mu\, P_{2\mu} = -
M_1^\mu\, q_{\perp\,\mu}/ y_q$, and, therefore, we obtain another
form of the impact factors:
 \be
  J_1 = -{\sqrt{2}\over sy_q} \,M_1^\mu\, q_{\perp\,\mu}\,,\;\; J_2 =
  -{\sqrt{2}\over sx_q} \,M_2^\nu\, q_{\perp\,\nu}\,.
  \label{25}
 \ee
This representation is important, since it shows that at small
transverse momentum of the exchanged photon the $J_{1,2}$ behave
as
 \be
  J_{1,2} \propto |{\bf q}_\perp| \;\;\mbox{at}\;\;{\bf q}_\perp \to 0\,.
  \label{26}
 \ee
 In our further analysis we will combine various contributions
of the impact factor into expressions which clearly exhibit such a
behaviour.

{\bf Vertices instead of spinor lines.} Let us consider a virtual
electron in the amplitude $M_1$ with
 $p=(E, {\bf p})$,  $E>0$ and virtuality
$p^2-m^2$. Due to jet kinematics, $|p^2-m^2| \ll E^2$. We
introduce an artificial energy
 $
  E_p = \sqrt{m^2+{\bf p}^2}
 $
and the bispinors  $u_{\bf p}^{(\lambda)}$ and $ v_{\bf
p}^{(\lambda)}$ corresponding to a real electron and a real
positron with 3-momentum $ {\bf p}$ and energy $E_p$. In the
high-energy limit $E-E_p = (p^2 -m^2)/(2E)$ and
 \be
  \hat p+m \approx  u_{\bf p}^{(\lambda)}
  \bar{u}_{\bf p}^{(\lambda)}+ {p^2
  -m^2\over 4 E^2} \, v_{-{\bf p}}^{(\lambda)} \bar{v}_{-{\bf
  p}}^{(\lambda)}\,.
  \label{30}
 \ee
Using this equation for all virtual electrons, we are able to
substitute the numerators of all spinor propagators by vertices
involving real electrons and real positrons. These generalized
vertices {\it are finite} in the limit $s\to \infty$. On the
contrary, a numerator like $\hat p+m$ is a sum of a finite term
$\hat p_\perp+m$ and an unpleasant combination $E\gamma^0 - p_z
\gamma_z$ of large terms that requires special care. Therefore,
those replacements significantly simplify all calculations.

More detailed considerations show that only three types of
vertices are needed to calculate the impact factors involving the
emission of real photons. The numerator of the spinor propagator
$\hat p+m$ in (\ref{30}) consists of two terms. The first term
corresponds to the simple replacements
 \be
\hat p+m \to  u_{\bf p}^{(\lambda)}\,\bar{u}_{\bf p}^{(\lambda)}
 \label{14a}
 \ee
and leads to the vertex for the transition
$e(p) + \gamma^*(q) \to e (p')$
(where $\gamma^*(q)$ is a virtual photon with
 energy fraction $x_q=0$ and an ``effective polarization vector'' $e_q
\equiv \sqrt{2} P_2/ s$)
  \be
  V(p) \equiv {V}_{\lambda \lambda'} (p) =
  \bar{u}_{{\bf p}'}^{(\lambda')}  \hat {e}_q \,{u}_{\bf
  p}^{(\lambda)}= \sqrt{2} {E'\over E_1} \delta_{\lambda
  \lambda'} \Phi
  \label{15a}
 \ee
and to the vertex for the transition $e(p) \to  e(p')+ \gamma(k)$
(where $\gamma(k)$ is a real photon with helicity $\Lambda$)
 \bea
\lefteqn{ V(p,\;k) = \bar{u}_{\bf p'}^{(\lambda')} \: \hat
{e}^{(\Lambda)\,*} \:
  {u}_{\bf p}^{(\lambda)}=
}
  \nn\\
\lefteqn{
=\bigg[\delta_{\lambda \lambda'}\,
  2\, \left( {e}^{(\Lambda)\,*} p \right)\, \left(1- x\,
  \delta_{\Lambda,- 2\lambda}\right) +
}
  \nn\\
\lefteqn{
  +  \delta_{\lambda,-
  \lambda'}\,\delta_{\Lambda, 2\lambda}\, \sqrt{2}\, mx
  \,\bigg]\, \Phi
}
 \eea
with $x=\omega/ E$ and
 \be
 \Phi = \sqrt{E\over E'}\, {\mathrm
  e}^{{\rm i} (\lambda' \varphi' - \lambda \varphi)},\;
  {e}p = {e}_\perp  \left(p_\perp - {k_\perp \over x} \right) .
\label{40}
 \ee

The second term in $\hat p+m$ of (\ref{30}) corresponds to the more
complicated replacement
 \be
\hat p+m \to  {p^2-m^2\over 4 E^2} \, v_{-{\bf p}}^{(\lambda)}
\bar{v}_{-{\bf p}}^{(\lambda)} \approx {p^2-m^2\over 4 EE_2}\,
{\hat P}_2\, .
  \label{18a}
 \ee
Since this expression contains a factor proportional to the
denominator of the spinor propagator, that denominator is
cancelled and a new vertex with four external lines (incoming and
outcoming leptons and two emitted photons) can be introduced:
 \bea
\lefteqn{
V(p,k_1,k_2)=
  }
  \\
 \lefteqn{  =
  {1\over 4(E-\omega_1)E_2} \bar{u}_{{\bf p}'}^{(\lambda')} \,
  \hat {e}^{(\Lambda_2)*}(k_2)\, \hat{P}_2 \,
  \hat{e}^{(\Lambda_1)*}(k_1)\, {u}_{{\bf p}}^{(\lambda)}
  }
\nn\\
 \lefteqn{
   =- 2{E'\over E-\omega_1}\,
  \delta_{\lambda \lambda'}\delta_{\Lambda_1, 2\lambda}
  \,\delta_{\Lambda_1,-\Lambda_2}
  \Phi
  \nn
 }
 \eea
with $\Phi$ defined in (\ref{40}). This vertex is similar to
a vertex with four external particles in scalar QED.

\section{IMPACT FACTOR FOR THE SINGLE BREMSSTRAHLUNG }

The impact factor for the single brems\-strahlung along the
direction of the first electron  corresponds to the virtual
Compton scattering $e(p_1)+ \gamma^*(q) \to e(p_3)+ \gamma(k)$ and
has the form
 \bea
\lefteqn{
  J_1 (e_{\lambda_1}+\gamma^* \to e_{\lambda_3}+ \gamma_\Lambda
)=4\pi\alpha \, \times
}
  \\
\lefteqn{
\bar u_3  \left( \frac{{\hat e}_q
  (\hat{p}_1-\hat k +m)\hat{e}^*}{2p_1k}
- \frac{\hat{e}^* (\hat{p}_3 +\hat k +m)
   {\hat e}_q}{2p_3k}\right) u_1\,.
  \nn
}
 \eea
Here for $\hat{p}_1-\hat k +m$ and $\hat{p}_3 +\hat k +m$ we can
use the simple substitution (\ref{14a}) that allows us to
eliminate the numerators of the two spinor propagators and to
introduce the vertices $V(p)$ and $V(p,k)$:
\bea
\label{1r}
\lefteqn{
J_1 = 4\pi \alpha  \, \times
}
\\
\lefteqn{
  \left[ {V(p_1,k) V(p_1-k)\over 2p_1k}\right.
- \left.{V(p_1)V(p_3+k,k)\over 2p_3k} \right].
}
 \nn
 \eea
This impact factor depends on the energy fractions $x=\omega/E_1$,
$X_3 = E_3/E_1$, and the transverse momenta ${\bf k}_\perp$, ${\bf
p}_{3\perp}$ of the final particles in the first jet with
the relations: $x+X_3=1$ and ${\bf k}_\perp+ {\bf p}_{3\perp}={\bf
q}_\perp$. In particular, the denominators in (\ref{1r}) are:
\bea
  \lefteqn{
  2p_1k= (m^2x^2+{\mathbf k}_{\perp}^2)/x \,,
  }
  \label{1ra}\\
  \lefteqn{
  2p_3k= [m^2x^2+({\bf k}_{\perp}-x{\bf q})^2)]/[x(1-x)] \,.
  }
  \nn
\eea

We further rearrange $J_1$ into an expression which clearly
exhibits the behaviour (\ref{26}). Using the simple relation
 \bea
    \label{3r}
\lefteqn{
V(p_3+k,k)= V(p_1 +q,k) =
}
   \\
\lefteqn{
V(p_1,k) +  2 \,
  \left(q_\perp\,{e}^{(\Lambda)\,*}_\perp \right)\, \left( 1- x\,
  \delta_{\Lambda,- 2\lambda_1} \right)\, \delta_{\lambda_1
  \lambda_3}\,,
}
\nn
  \eea
we immediately obtain the final result
  \bea
\lefteqn{
J_1 = \sqrt{2}\, 4\pi \alpha \bigg\{\left[ \delta_{\lambda_1
\lambda_3}\, 2\,\left(  {e}^{(\Lambda)\,*} p_1
\right)\times\,\right.
}
 \nn\\
\lefteqn{
 \left. \left(1- x\, \delta_{\Lambda,- 2\lambda_1}\right)
+ \delta_{\lambda_1,- \lambda_3}\,\delta_{\Lambda, 2\lambda_1}\,
\sqrt{2}\, m x\right] A_1
}
 \nn\\
\lefteqn{
 +q_\perp B_1\bigg\} \, \Phi_{13}\,,
}
   \label{4r}
  \eea
where
  \bea
\lefteqn{
    A_1={1-x\over 2p_1k} -{1\over 2p_3k}\,,
 \;\;
\Phi_{13} = {1\over \sqrt{X_3}} \, {\mathrm e}^{{\rm
i}(\lambda_3
  \varphi_3 -\lambda_1 \varphi_1)}\,,
}
 \nn\\
\lefteqn{
    B_1 =-{{e}^{(\Lambda)\,*}_\perp \over p_3k}\, \left( 1- x\,
  \delta_{\Lambda,- 2\lambda_1} \right)\, \delta_{\lambda_1
  \lambda_3} \,.
}
  \label{6r}
  \eea
The impact factor  (\ref{4r}) is a simple and compact expression
for all 8 helicity states written in a form that all individual
large (compared to $ q_\perp$) contributions are cancelled.
Indeed, the last term in $J_1$ is directly proportional to
$q_\perp$ and $A_1 \propto q_\perp$ due to (\ref{1ra}).

\section{IMPACT FACTOR FOR THE DOUBLE BREMSSTRAHLUNG}

\begin{figure}[!htb]
  \begin{center}
  \unitlength=2.00mm
  \begin{picture}(35.00,16.00)(0.00,0.00)
    \put(1.00, 8.50){\line(1,0){9.0}}
    \put(1.00,8.50){\vector(1,0){1.50}}
    \put(4.66,8.50){\vector(1,0){2.90}}
    \put(7.00,8.50){\line(1,0){9.0}}
    \put(7.00,8.50){\vector(1,0){8.0}}
    \put(10.40,8.50){\vector(1,0){1.0}}
    \put(2.00,7.00){\makebox(0,0)[cc]{$p_1$}}
    \put(15.00,7.00){\makebox(0,0)[cc]{$p_3$}}
    \put(13.00, 2.50){\line(0,1){2.00}}
    \put(13.00,5.00){\line(0,1){2.00}}
    \put(13.00,7.50){\line(0,1){1.00}}
    \put(13.00,5.00){\vector(0,1){2.00}}
    \put(11.00,5.50){\makebox(0,0)[cc]{$q$}}
    \put(14.00,13.50){\line(1,1){1.8}}
    \put(11.50,11.00){\line(1,1){1.8}}
    \put(9.00,8.50){\line(1,1){1.8}}
    \put(11.50,11.00){\vector(1,1){1.8}}
    \put(9.00,11.20){\makebox(0,0)[cc]{$k_2$}}
    \put(9.00,13.50){\line(1,1){1.8}}
    \put(6.50,11.00){\line(1,1){1.8}}
    \put(4.00,8.50){\line(1,1){1.8}}
    \put(6.50,11.00){\vector(1,1){1.8}}
    \put(4.00,11.20){\makebox(0,0)[cc]{$k_1$}}
    \put(17.00, 8.50){\line(1,0){9.0}}
    \put(17.00,8.50){\vector(1,0){1.50}}
    \put(20.66,8.50){\vector(1,0){2.90}}
    \put(23.00,8.50){\line(1,0){11.0}}
    \put(23.00,8.50){\vector(1,0){9.0}}
    \put(26.40,8.50){\vector(1,0){1.0}}
    \put(25.00, 2.50){\line(0,1){2.00}}
    \put(25.00,5.00){\line(0,1){2.00}}
    \put(25.00,7.50){\line(0,1){1.00}}
    \put(25.00,5.00){\vector(0,1){2.00}}
    \put(23.00,5.50){\makebox(0,0)[cc]{$q$}}
    \put(25.00,13.50){\line(1,1){1.8}}
    \put(22.50,11.00){\line(1,1){1.8}}
    \put(20.00,8.50){\line(1,1){1.8}}
    \put(22.50,11.00){\vector(1,1){1.8}}
    \put(20.00,11.20){\makebox(0,0)[cc]{$k_1$}}
    \put(34.00,13.50){\line(1,1){1.8}}
    \put(31.50,11.00){\line(1,1){1.8}}
    \put(29.00,8.50){\line(1,1){1.8}}
    \put(31.50,11.00){\vector(1,1){1.8}}
    \put(29.00,11.20){\makebox(0,0)[cc]{$k_2$}}
  \end{picture}
  \begin{picture}(35.00,16.00)(0.00,0.00)
   \put(10.00, 8.50){\line(1,0){9.0}}
    \put(10.00,8.50){\vector(1,0){1.50}}
    \put(13.66,8.50){\vector(1,0){2.90}}
    \put(16.00,8.50){\line(1,0){11.0}}
    \put(16.00,8.50){\vector(1,0){9.0}}
    \put(21.40,8.50){\vector(1,0){1.0}}
    \put(13.00, 2.50){\line(0,1){2.00}}
    \put(13.00,5.00){\line(0,1){2.00}}
    \put(13.00,7.50){\line(0,1){1.00}}
    \put(13.00,5.00){\vector(0,1){2.00}}
    \put(11.00,5.50){\makebox(0,0)[cc]{$q$}}
    \put(23.00,13.50){\line(1,1){1.8}}
    \put(20.50,11.00){\line(1,1){1.8}}
    \put(18.00,8.50){\line(1,1){1.8}}
    \put(20.50,11.00){\vector(1,1){1.8}}
    \put(18.00,11.20){\makebox(0,0)[cc]{$k_1$}}
    \put(28.00,13.50){\line(1,1){1.8}}
    \put(25.50,11.00){\line(1,1){1.8}}
    \put(23.00,8.50){\line(1,1){1.8}}
    \put(25.50,11.00){\vector(1,1){1.8}}
    \put(23.50,11.20){\makebox(0,0)[cc]{$k_2$}}
  \end{picture}
  \end{center}
  \vspace{-10mm}
  \caption{Feynman diagrams for the impact factor related to the
  double bremsstrahlung, diagrams with $k_1 \leftrightarrow k_2$
  photon exchange have to be added.}
  \label{fig:13}
\end{figure}
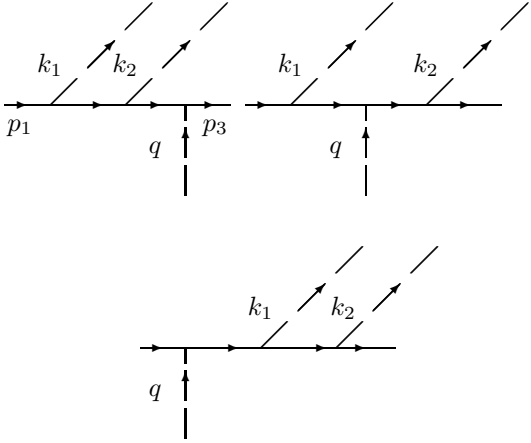

The impact factor  $J_1$ for the double brems\-strahlung along the direction of
the first electron $e(p_1)+ \gamma^*(q) \to e(p_3)+ \gamma(k_1)+
\gamma(k_2)$ corresponds to six diagrams, three of them are shown
in Fig.~\ref{fig:13}.

{\bf  Notations.}
$J_{1}$ depends  only on the
energy fractions
$x_{1,2} =\omega_{1,2}/E_1$, $X_3 = E_3/E_1$ with $x_1+x_2+X_3=1$
and on the transverse momenta of the final particles in the
combinations:
 \bea
\lefteqn{
  q_\perp = k_{1\perp} +  k_{2\perp} +p_{3\perp}, \;
  r_j= X_3 k_{j\perp} - x_j p_{3\perp},
}
\\
\lefteqn{
   K_j = k_{jx} + {\rm i}\, k_{jy}, \; Q = q_{x} + {\rm i}\, q_{y},
  \; R_j = r_{jx} + {\rm i}\, r_{jy}.
}
\nn
\eea
The denominators of the spinor propagators are:
  \bea
\lefteqn{
   a_j\equiv-(p_1-k_j)^2+m^2=
   \frac{1}{x_j}(m^2x_i^2+{\mathbf k}_{j\perp}^2)\,,
}
  \nn
  \\
\lefteqn{
  b_j \equiv (p_3+k_j)^2-m^2=\frac{1}{x_j X_3}(m^2 x_j^2+
  {\mathbf r}_{j\perp}^2)\,,
}
  \nn
  \\
\lefteqn{
  a_{12}=a_{21} \equiv -(p_1-k_1-k_2)^2+m^2=
}
  \\
\lefteqn{
  = a_1+ a_2 -
  \frac{1}{x_1x_2}
  \left(x_1{\mathbf{k}}_{2\perp}
  -x_2{\mathbf{k}}_{1\perp} \right)^2\,,
}
\nn
\\
\lefteqn{
 b_{12}=b_{21}\equiv (p_3+k_1+k_2)^2-m^2=
}
  \nn\\
\lefteqn{
=b_1+
  b_2+\frac{1}{x_1x_2}
  \left(x_1{\mathbf{k}}_{2\perp}
  -x_2{\mathbf{k}}_{1\perp} \right)^2\,.
}
  \nn
 \eea

{\bf General formula.} To calculate the impact factor  $J_1$, we
follow along the electron line from left to right in the diagrams of
Fig.~\ref{fig:13} and write down the corresponding vertices:
   \bea
\lefteqn{
 {J_{1}\over (4\pi\alpha)^{3/2}} = \frac{V(p_1, k_1)
  V(p_1-k_1, k_2)V(p_3-q)}{a_1a_{12}}
}
  \nn\\
\lefteqn{
    -\frac{V(p_1, k_1) \, V(p_1-k_1)\,  V(p_1-k_1+q, k_2)}{a_1 b_2}
}
 \nn\\
\lefteqn{
   + \frac{V(p_1)\, V(p_1 +q, k_1)\, V(p_1-k_1+q, k_2)}{b_{12} b_2}
}
  \nn\\
\lefteqn{
    -\frac{V(p_1, k_1,k_2) \,  V(p_3-q)}{a_{12}}\
}
  \nn\\
 \lefteqn{
    + \frac{ V(p_1)\,  V(p_1+q,k_1, k_2)} {b_{12}}
}
    \nn\\
\lefteqn{
+ (k_1 \leftrightarrow k_2)\,.
}
    \eea
Then we present $J_1$ in the form
 \bea
    \label{imp}
\lefteqn{
 J_{1}= \sqrt{2} \, (4\pi\alpha)^{3/2}\, X_3 \,
 \left(1 +{\cal P}_{12}\right)\,\times
}
  \\
\lefteqn{
  M_{\lambda_1\, \lambda_3}^{\Lambda_1\, \Lambda_2}\, (x_1, x_2,
   k_{1\perp},  k_{2\perp}, p_{3\perp}) \, \Phi_{13}\,,
}
   \nn
   \eea
where we introduce the permutation operator
\begin{eqnarray}
  {\cal{P}}_{12}f(k_1,e_1;k_2,e_2)= f(k_2,e_2;k_1,e_1)
\end{eqnarray}
and the factor $\Phi_{13}$ from (\ref{6r}) which includes the
common phase. This allows us to omit below all factors $\Phi$ from
the vertices $V(p)$, $ V(p, k)$ and $V(p,k_1, k_2)$.

Using relations similar to (\ref{3r}), we transform $M$ into
an expression which clearly exhibits the behaviour (\ref{26}):
 \bea
  \label{beh}
\lefteqn{
  X_3 M_{\lambda_1\,\lambda_3}^{\Lambda_1 \Lambda_2}=
}
\nn \\
\lefteqn{
  +  A_2 \, V_{\lambda_1\lambda}^{\Lambda_1}(p_1,k_1)
   \,
   V_{\lambda\lambda_3}^{\Lambda_2}(p_1-k_1 +q, k_2) +
}
\\
\lefteqn{
+
q_\perp {B_2}_{\lambda_1\,\lambda_3}^{\Lambda_1 \Lambda_2} +
  {\tilde A}_2\, V_{\lambda_1\lambda_3}^{\Lambda_1 \Lambda_2}(p_1, k_1,k_2)\,,
}
  \nn
  \eea
 where
 \bea
\lefteqn{
A_2=  \frac{X_3}{a_1a_{12}} - \frac{1-x_1}{a_1b_2}+
  \frac{1}{b_{12} b_2}\,,
}
  \nn\\
\lefteqn{
  {\tilde A}_2 =-\frac{X_3}{a_{12}} + \frac{1}{b_{12}}
}
  \eea
and the transverse 4--vector $B_2$ is
 \bea
 \lefteqn{
   {B_2}_{\lambda_1\,\lambda_3}^{\Lambda_1 \Lambda_2} =
  - X_3\,{2 {e}^{(\Lambda_2)\,*}_\perp\over a_1a_{12}}\,
  V_{\lambda_1 \lambda_3}^{\Lambda_1}(p_1,k_1)\, \times
}
  \nn\\
\lefteqn{
  \left( 1- {x_2\over 1-x_1}\, \delta_{\Lambda_2,- 2\lambda_3}
  \right) +
}
 \\
\lefteqn{
  +{2 {e}^{(\Lambda_1)\,*}_\perp\over  b_{12} b_2}\,
  V_{\lambda_1 \lambda_3}^{\Lambda_2}(p_1-k_1+q, k_2)
}
   \nn\\
\lefteqn{
   \times \left( 1- x_1\,
  \delta_{\Lambda_1,- 2\lambda_1} \right)\,.
}
\nn
     \eea
Now it is not difficult to check that $A_2\propto q_\perp$,
${\tilde A}_2 \propto q_\perp$ and, therefore, $J_1 \propto
q_\perp$ as well.

Again equations (\ref{imp}) and (\ref{beh}) represent a very simple and
compact expression for all 16 helicity states, where all
individual large (compared to $ q_\perp$) contributions have been
rearranged into finite expressions.

{\bf Explicit expressions.} To find the amplitudes with given
initial and final helicities, it is sufficient to substitute in
the above equations the expressions for the vertices.
As a result, we find:
 \bea
\lefteqn{
 M_{++}^{++}=
}
\nn\\
\lefteqn{
  2 \left\{
  A_2 \frac{K_1^* R_2^*}{x_1 x_2 X_3}
  +\frac{K_1^* Q^*}{x_1 a_1 a_{12} }
  - \frac{Q^* R_2^*}{x_2 X_3 b_{12} b_2 } \right\}\,,
}
 \nn\\
\lefteqn{
  M_{++}^{--}= X_3 \left(M_{++}^{++} \right)^*\,,
}
  \nn \\
\lefteqn{
   M_{++}^{-+}=-2 (1-x_1)
}
   \nn\\
\lefteqn{
 \times\left(
  A_2 \frac{K_1 R_2^*}{x_1x_2X_3}
  +\frac{K_1 Q^*}{x_1 a_1 a_{12} }
  -\frac{Q R_2^*}{x_2 X_3  b_{12} b_2 }\right) \,,
}
  \nn \\
\lefteqn{
 M_{++}^{+-}= {X_3\over (1-x_1)^2} \left(M_{++}^{-+}\right)^*+
}
  \\
\lefteqn{
 +  \frac{2}{1-x_1} \,\left( m^2 A_2 \frac{x_1 x_2}{X_3} -
  \tilde{A}_2\right)\,,
}
  \nn\\
\lefteqn{
  M_{+-}^{+-}= 2 m x_1 \left( A_2 \frac{R_2}{x_2 X_3}
  +\frac{Q}{a_1 a_{12}}\right)\,,
}
  \nn\\
\lefteqn{
  M_{+-}^{-+}= 2 m \frac{x_2}{X_3}  \left( A_2 \frac{K_1}{x_1} -
  \frac{Q}{b_{12}   b_2}\right)\,,
}
  \nn\\
\lefteqn{
 M_{+-}^{--}=0\,,
}
  \nn\\
\lefteqn{
 M_{+-}^{++}= -{1\over 1-x_1}\left(X_3\,M_{+-}^{+-}+
  M_{+-}^{-+}\right)^*\,.
}
  \nn
 \eea
Other amplitudes can be found due to the parity conservation
relation
 \be
M_{-\lambda_1\;-\lambda_3}^{-\Lambda_1\, -\Lambda_2}= -\,
(-1)^{\lambda_1+\lambda_3}\,\left(M_{\lambda_1\;\lambda_3}^{\Lambda_1
\Lambda_2}\right)^*\,.
 \ee

The generalization of the results obtained for the single and
double bremsstrahlung to the bremsstrahlung of $n$ photons can be
done straightforwardly. To demonstrate this, we have considered in detail
the case $n=3$ in \cite{CSS1}.

\section{SUMMARY}

{\bf 1.} We have formulated a new effective method to calculate
all helicity amplitudes for bremsstrahlung jet-like QED processes
at tree level. The main advantage of our method consists in using
simple universal ``building blocks'' --- transition vertices with
real leptons. Those vertices replace efficiently the spinor
structure involving leptons of small virtuality in the impact
factors, making the calculations short and transparent for any
final helicity state.

{\bf 2.} In the considered case of bremsstrahlung we have found
that only three nonzero transition vertices are required. The
properties of these vertices determine all nontrivial general
properties of the helicity amplitudes (for details
see~\cite{CSS1,CSS2}).

{\bf 3.} We have outlined in this contribution how to calculate
the impact factors for single and double bremsstrahlung. The case
of triple bremsstrahlung can be found in~\cite{CSS1}. It allows us
to give a complete analytic and compact description of all
helicity amplitudes in $e^-e^\pm$ scattering with the emission of
up to three photons along the directions of each initial leptons
(in that case $2^5 \times 2^5=1024$ different helicity amplitudes
are involved). The corresponding calculations for the case of
lepton pair production can be found in~\cite{CSS2}.

{\bf 4.} Since by construction individual large (compared to
$q_\perp$) contributions have been rearranged into finite
expressions, the formulae obtained for the amplitudes are very
convenient for numerical calculations of various cross sections.


\begin{thebibliography}{99}

\bibitem{CSS1}
C.~Carimalo, A.~Schiller, V.G.~Serbo, Eur. Phys. J. C  23 (2002)
633.

\bibitem{CSS2}
  C.~Carimalo, A.~Schiller, V.G.~Serbo, hep-ph/0303257,
  to appear in Eur. Phys. J. C (2003).

\bibitem{kuraevnp85}
  E.A.~Kuraev, A.~Schiller, V.G.~Serbo, Nucl. Phys. B 256 (1985)
  189.

  \bibitem{mesons}
  I.F.~Ginzburg, S.L.~Panfil, V.G.~Serbo, Nucl. Phys. B 284 (1987)
  685, ibid.  B 296 (1988) 581.

 \bibitem{BGMS}
  V.M.~Budnev, I.F.~Ginzburg, G.V.~Meledin, V.G.~Serbo, Phys. Rep.
   15 C (1975) 181.

\bibitem{BFKK}
  V.N.~Baier, V.S.~Fadin, V.A.~Khoze, E.A.~Kuraev, Phys. Rep. 78
  (1981) 293.

  \bibitem{KSSSh00}
  E.A.~Kuraev, A.~Schiller, V.G.~Serbo, B.G.~Shaikhatdenov, Nucl.
  Phys. B 570 (2000) 359.

\bibitem{Kessler60}
  P.~Kessler, Nuovo Cim. 17 (1960) 809.

\bibitem{BFKhquasi}
  V.M.~Baier, V.S.~Fadin, V.A.~Khoze, Nucl. Phys. B 65 (1973) 381.


\end{thebibliography}
\end{document}